\documentclass[aps,prb,floatfix,reprint]{revtex4-1}

\usepackage{graphicx}
\usepackage{subfigure}
\usepackage{amsfonts}
\usepackage{amsmath}
\usepackage{amssymb}
\usepackage{xcolor}

\begin{document}
\title{Entropy current and efficiency of quantum machines driven by nonequilibrium incoherent reservoirs}

\author{Sebastian E. Deghi$^{1}$ and Ra\'ul A. Bustos-Mar\'un$^{1,2,*}$}
\affiliation{$^{1}$ Instituto de F\'isica Enrique Gaviola (CONICET) and FaMAF (Universidad Nacional de C\'ordoba), Ciudad Universitaria, C\'ordoba 5000, Argentina.
$^{2}$ Facultad de Ciencias Qu\'imicas (Universidad Nacional de C\'ordoba), Ciudad Universitaria, C\'ordoba 5000, Argentina.
}
\email{rbustos@famaf.unc.edu.ar}

\begin{abstract}
Nanotechnology has not only provided us the possibility of developing quantum machines but also noncanonical power sources able to drive them.
Here we focus on studying the performance of quantum machines driven by arbitrary combinations of equilibrium reservoirs and a form of engineered reservoirs consisting of noninteracting particles but whose distribution functions are nonthermal.
We provide the expressions for calculating the maximum efficiency of those machines without needing any knowledge of how the nonequilibrium reservoirs were actually made.
The formulas require the calculation of a quantity that we term entropy current, which we also derive.
We illustrate our methodology through a solvable toy model where heat ``spontaneously'' flows against the temperature gradient.
\end{abstract}

\maketitle

\section{Introduction}

The tendency toward miniaturization reached nanoscale a long time ago.
This opened up the door to the design and control of different forms of quantum machines, such as quantum motors, quantum pumps, quantum heat engines, or quantum heat pumps ~\cite{brouwer1998,dundas2009,bustos2013,bohrbrask2015,esposito2015,lu2016,ludovico2016,ludovico2016B,calvo2017,benenti2017,romeo2018,whitney2018,bustos2019,lin2019,zimbovskaya2020}.
These systems have been extensively studied during past years, including their dynamical and thermodynamical aspects.
However, paraphrasing Feynman, there is still plenty of room at the bottom, and new proposals keep surprising us.
The possibility of using noncanonical power sources, such as nonequilibrium reservoirs~\cite{scully2003,alicki2015,manzano2016,gianluca2017,ghosh2018,assis2019,sanchez2019} or Maxwell's demons
~\cite{koski2014,camati2016,vidrighin2016,chida2017,masuyama2018,strasberg2013},
is a tantalizing new direction which may not only offer alternative ways of controlling quantum machines, but also sheds light on the thermodynamics of real and thought experiments.

In the literature, there are different forms of demonlike ``engineered reservoirs '', some of them involving subtle quantum coherences or correlation effects~\cite{scully2003,manzano2016,gianluca2017,ghosh2018}.
Here we focus on a somewhat simpler kind of engineered reservoir that we call nonequilibrium incoherent reservoirs (NIRs)~\footnote{We include the word incoherent in the name ``nonequilibrium incoherent reservoirs'' to emphasis the difference with other kind of reservoirs with engineered quantum coherences, see Refs.~\cite{scully2003,manzano2016,gianluca2017,ghosh2018}}.
These reservoirs consist of noninteracting quantum particles (just as the usual ones in quantum transport~\cite{ludovico2016,ludovico2016B,benenti2017,whitney2018,bustos2019}), but with distribution functions that are nonthermal.
In the context of quantum transport, nonthermal distributions in mesoscopic systems have not only been studied theoretically~\cite{karzig2010,kovrizhin2012,ajisaka2013,alicki2015,stegmann2018} but also experimentally observed, e.g., in mesoscopic wires~\cite{pothier1997}, carbon nanotubes~\cite{chen2009,bronn2013}, quantum Hall edge channels~\cite{altimiras2010} and graphene~\cite{voutilainen2011}.

Despite being simpler than other proposals, NIRs can give rise to fascinating phenomena.
For example, it has been shown that NIRs can act as a sort of Maxwell's demon
that, without injecting energy or particles into a device, allows it, e.g., to pump heat against a temperature gradient~\cite{sanchez2019}.
This may have important applications, as the prospect of a nanorefrigerator that works without having to inject energy into it, which at some point should dissipate as heat~\cite{bustos2019}, seems ideal.
Beyond this proposal, other forms of quantum devices, driven by more general combinations of equilibrium and nonequilibrium reservoirs, are also possible and interesting to study.
However, there is not a general thermodynamics description of this broad class of devices. 
Therefore, to calculate, e.g., the efficiency, one usually needs to know how the NIRs were made, starting from equilibrium reservoirs.
Here we discuss the thermodynamics and the efficiency of this class of devices but from a description that only requires the probability distribution function of the NIRs.
Our formulation is based on the calculation of a quantity that we dubbed entropy current, which here is derived within a semiclassical approach. See also Appendix~\ref{app:T_eff} for an alternative derivation based on von Neumann entropy.
\begin{figure}
\includegraphics[width=2.8in]{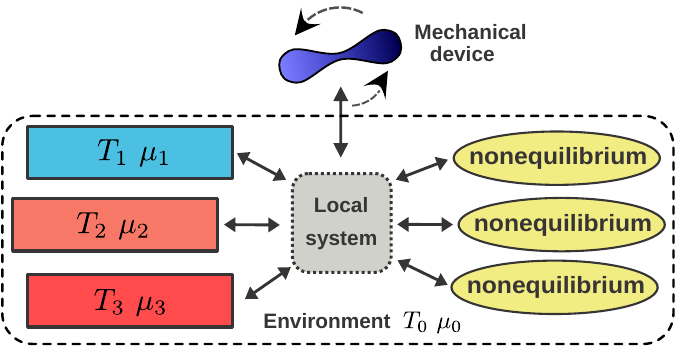}
\caption{
 General scheme of the type of system treated. A local system, connected (or not) to a mechanical device, interchanges particles with equilibrium reservoirs (at temperature $T_{i}$ and chemical potential $\mu_{i}$) and nonequilibrium incoherent reservoirs.
}
\label{fig:1}
\end{figure}

\section{Entropy current}
We start by considering a reservoir with total energy $U$ and a large number, $N$, of indistinguishable noninteracting particles.
Let us divide its spectrum into groups of levels called cells and separated by an energy interval
$\delta \epsilon$. The number of states of the $i$-cell is $g_{i}$, the average energy of the cell is $\varepsilon_i$, and $n_{i}$ is the number of particles occupying states within the $i$-cell for a particular configuration of the reservoir.
The entropy $S$ of this reservoir can be calculated from~\cite{huang1987}
\begin{eqnarray}
S & = & k_{B}\ln \left ( \sum_j W\{n_{i}\}_j \right ) ,
\end{eqnarray}
where, $k_{B}$ is the Boltzmann constant, $W\{n_{i}\}_j$ is the number of states of the reservoir corresponding to a particular set $j$ of occupations $\{n_{i}\}$, and the summation runs over all sets of occupations compatible with the total energy $U$ and the number of particles $N$ of the reservoir.
Importantly, we are assuming that fluctuations around a given value of $n_{i}$ are negligible, and thus we can consider it fixed. Therefore, $\sum_j W\{n_{i}\}_j \approx W\{n_i\}_0$ where $\{n_i\}_0$ is the set of occupation numbers fixed, either by an external agent or by the maximization of the entropy of the reservoir.

As we are dealing with noninteracting particles, we have $W\{n_{i}\}_0=\prod_{i}w_{i}$, where $w_{i}$ denotes the number of ways in which $n_{i}$ particles can be assigned to the $i$-cell of the reservoir with $g_{i}$ states.
Then, for bosons and fermions, the following relation holds~\cite{huang1987}
\begin{eqnarray}
w_{i}^{(\mathrm{bosons})} & = & \left(n_{i}+g_{i}-1\right)!/\left[n_{i}!\left(g_{i}-1\right)!\right],\\
w_{i}^{(\mathrm{fermions})} & = & g_{i}!/\left[n_{i}!\left(g_{i}-n_{i}\right)!\right]
\end{eqnarray}
and hence
\begin{eqnarray}
\frac{\ln w_{i}}{g_{i}} & \approx & \overline{n}_{i}\left(\ln\left[\frac{1}{\overline{n}_{i}}\pm1\right]\pm\frac{1}{\overline{n}_{i}}\ln\left[1\pm\overline{n}_{i}\right]\right),\label{eq:lnw}
\end{eqnarray}
where $\overline{n}_{i}=n_{i}/g_{i}$ is the average occupation, the $+$ sign is for bosons and the $-$ sign is for fermions.
Notice that we assumed $g_{i}\gg1$ for bosons and used the Stirling approximation.

With the aid of Eq. \ref{eq:lnw}, one can calculate the entropy of a reservoir $\alpha$ by $S_{\alpha} = k_{B} \sum_i \ln w_{i\alpha}$.
But now, suppose there is a flux of particles per unit energy $\dot{n}_\alpha\left(\varepsilon_i\right)$ entering the reservoir, where $\dot{n}_\alpha\left(\varepsilon_i\right) = \dot{\overline{n}}_{i \alpha} (g_{i \alpha}/ \delta \epsilon) $ and we assume the spectrum of the reservoir remains constant.
We want to know the change per unit time of the entropy as a consequence of that, from now on the ``entropy current''. This is given by
\begin{equation}
 \dot S_{\alpha} =
 k_{B} \sum_i
 \frac{
 \partial \ln w_{i\alpha}
 }
 {
  \partial \overline{n}_{i\alpha}
  }
  \frac{\dot{n}_{\alpha}\left ( \varepsilon_i \right )}{g_{i\alpha}} \delta \epsilon ,
\end{equation}
where the dot stands for time-derivative.
Deriving $\ln w_{i\alpha}$ and turning the summation into an integration, yields
\begin{eqnarray}
\dot{S}_{\alpha} & = & k_{B}
\int
\dot{n}_\alpha\left(\varepsilon_i\right) \ln\left(\frac{1}{ f_\alpha\left(\varepsilon\right)}\pm1\right)
\mathrm{d} \varepsilon.
\label{eq:dotS}
\end{eqnarray}
Here we replace $\overline{n}_\alpha \left(\varepsilon_i \right)$ by $f_\alpha\left(\varepsilon\right)$ to emphasis the fact that we are considering large reservoirs. There, the number of states $g_{i\alpha}$ within the energy interval $\delta \epsilon$ is so huge that one can consider $\overline{n}_\alpha \left(\varepsilon_i \right)$ time-independent.
Note, that for equilibrium reservoirs ($f_{\alpha}\left(\varepsilon\right)$ given by the Fermi-Dirac or the Bose-Einstein distribution functions at temperature $T_r$), one recovers the well known formula $\dot{S}_r=J_r/T_{r}$~\cite{ludovico2016,ludovico2016B,benenti2017,whitney2018,bustos2019}, where $J_r$ is the heat current
\begin{equation}
 \dot {S}_r= \frac{1}{T_r} \intop_{-\infty}^{\infty}\dot{n}\left(\varepsilon\right)\left(\varepsilon-\varepsilon_{F}\right)d\varepsilon .
\end{equation}
The entropy current can also be expressed in terms of an energy-dependent effective temperature, see Appendix~\ref{app:T_eff}.

\section{Efficiency of quantum machines  driven by NIRs}
We will consider a local system connected to a certain number of reservoirs at equilibrium,  denoted by $r$, but also connected to several NIRs, denoted by $l$, see Fig. \ref{fig:1}.
To add generality, we will also include the possibility that the local system is connected to a mechanical device. Thus, current-induced forces (or the possible external forces) should be taken into account~\cite{bustos2013,ludovico2016,ludovico2016B,calvo2017,benenti2017,whitney2018,bustos2019}.
Finally, we will assume fermionic reservoirs for the derivations. However, the results can be translated readily for the bosonic case or for mixed reservoirs (once particle conservation is appropriately taken into account in the latter).

We start by writing the time-derivative of the total energy $U_{\mathrm{total}}$ of the reservoirs
\begin{eqnarray}
\dot{U}_{\mathrm{total}} & = & \sum_{r}\left ( T_{r}\dot{S}_{r}+\dot{n}_{r}\mu_{r} \right ) + \sum_l \dot U_l \label{eq:sumU_r}
\end{eqnarray}
where $U_l$ is the internal energy of the reservoir $l$.
Equilibrium reservoirs are characterized by a temperature $T_{r}$, an entropy $S_{r}$, a number of particles $n_{r}$, and a chemical potential $\mu_{r}$.
For them, we used $\dot U_r=J_{r} +\dot{n}_{r}\mu_{r}$, where we identified the heat current $J_{r}$ with $T_{r}\dot{S}_{r}$.
Now we define $\delta T_{r}=T_{r}-T_{0}$ and $\delta\mu_{r}=\mu_{r}-\mu_{0}$, where $T_{0}$ and $\mu_{0}$ are just reference temperatures and chemical potentials but, for convenience, we will identify them with the temperature and chemical potential of the surrounding environment, see Fig. \ref{fig:1}.
Energy conservation imposes
\begin{equation}
\sum_{\alpha=r,l} (\dot{U}_{\alpha}+ \dot{U}_{\mathrm{s},\alpha}) +\dot{U}_{\mathrm{s}}  =-\mathcal{\dot{W}}_{F} , 
\end{equation}
where $\mathcal{\dot{W}}_{F}$ is the power delivered by current-induced forces (or the external ones) that might be acting on the local system, $\dot{U}_{\mathrm{s}}$ is energy current of the local system, and $\sum_{\alpha=r,l}\dot{U}_{\mathrm{s},\alpha}$ is the time-derivative of the couplings between the local system and the reservoirs.
Particle conservation imposes 
\begin{equation}
 \sum_{\alpha=r,l}\dot{n}_{\alpha}=-\dot{n}_{\mathrm{s}} ,
\end{equation}
where $\dot{n}_{\mathrm{s}}$ is the particle current of the local system.
The time-derivative of the entropy of all  reservoirs, $\dot{S}_{T}=\sum_{\alpha=r,l}\dot{S}_{\alpha}$, is obtained by using all the above on Eq. \ref{eq:sumU_r}, see ,e.g., Refs. ~\cite{ludovico2016,ludovico2016B,benenti2017,whitney2018,bustos2019}. The quantity $\dot{S}_{T}$ can be divided into a reversible ($\dot{S}^{\mathrm{(rev)}}_{\mathrm{total}}$) and an irreversible ($\dot{S}^{\mathrm{(irrev)}}_{T}$) components.
The results for $\dot{S}^{\mathrm{(irrev)}}_{T}$, the rate of entropy production, is
\begin{eqnarray}
T_{0}\dot{S}^{\mathrm{(irrev)}}_{T}
& = &
-\mathcal{\dot{W}}_{F}-\sum_{r}\left ( \dot{n}_{r}\delta\mu_{r}+J_{r} \frac{\delta T_{r}}{T_{r}} \right )
\notag \\ &&
- \sum_l \left ( \dot U_l - \mu_0 \dot n_l - T_0 \dot S_l  \right ),
\label{eq:dotS-1}
\end{eqnarray}
while the reversible component of $\dot{S}_T$ is given by
\begin{eqnarray}
T_{0}\dot{S}^{\mathrm{(rev)}}_{T} =
-\left(\sum_{\alpha=r,N}\dot{U}_{\mathrm{s},\alpha}+\dot{U}_{\mathrm{s}}-\mu_{0} \dot{n}_{\mathrm{s}} \right)
.
\end{eqnarray}
The second law of thermodynamics imposes $\dot{S}^{\mathrm{irrev}}_{T}\geq0$ .
Hence, after integrating Eq. \ref{eq:dotS-1} over a cycle of a quantum machine (with period $\tau$) working within the local system, the following inequality holds
\begin{eqnarray}
0 & \geq & \frac{\mathcal{W}_{F}+\sum_l \Delta\mathcal{E}_{l}}{\tau}
+\sum_{r} \left\langle \dot{n}_{r}\right\rangle \delta\mu_{r}
+ \left\langle J_{r}\right\rangle \left(\frac{\delta T_{r}}{T_{r}}\right).\label{eq:rop}
\end{eqnarray}
Here, $\mathcal{W}_{F}$ is the work per cycle done by the current-induced forces, $\left\langle \bullet \right\rangle =\intop_{0}^{\tau} \bullet dt/\tau$, and we define the quantity
\begin{equation}
\Delta\mathcal{E}_{l} = \Delta U_l - \mu_0 \Delta n_l - T_0 \Delta S_l . \label{eq:DeltaE}
\end{equation}
To understand the physical meaning of the term $\Delta\mathcal{E}_{l}$, instead of transforming the NIR from an initial to a final state, let us imagine the following processes: 1) Turn the initial state of the NIR into an equilibrium reservoir with $T_l$ and $\mu_l$ while keeping constant $U_l$ and $n_l$. 2) Change the energy and the number of particles in an amount of $\Delta U_l$ and $\Delta n_l$, respectively. 3) Turn the final state of the ``equilibrium'' $l$-reservoir into the desired final state of the NIR while keeping constant $U_l$ and $n_l$. The change of the energy of the ``equilibrium'' $l$-reservoir during step 2 is $\Delta S_l^{\mathrm{eq}} \delta T_l+\Delta n_l \delta \mu_l$. The minimum heat absorbed during steps 1 and 3 by the environment (at temperature $T_0$) to transform back and forth the nonequilibrium $l$-reservoir into its ``equilibrium'' counterpart is $T_0 (\Delta S^{\mathrm{eq}}_l-\Delta S_l)$. The sum of these two contributions is $\Delta\mathcal{E}$, see also Appendix~\ref{app:DE}.
Finally, note that $\mathcal{E}_l$ is like a grand potential but for NIRs, where $T_0$ and $\mu_0$ are used as the temperature and chemical potential of the reservoir~\cite{esposito2015}.

Equation \ref{eq:rop} sets the limits of the efficiency of a broad class of quantum machines.
For example, for adiabatic quantum motors and adiabatic quantum pumps Eq. \ref{eq:rop} gives, respectively
\begin{eqnarray}
1 & \geq & \frac{\mathcal{W}_{F}}{-\sum_{r}\tau\left\langle \dot{n}_{r}\right\rangle \delta\mu_{r}-\sum_l \Delta\mathcal{E}_{l}}\qquad\mathrm{and} \notag \\
1 & \geq & \frac{\sum_{r}\tau\left\langle \dot{n}_{r}\right\rangle \delta\mu_{r}}{-\mathcal{W}_{F}-\sum_l \Delta\mathcal{E}_{l}}. \label{eq:eta_AQM-AQP}
\end{eqnarray}
For adiabatic quantum motors, $\mathcal{W}_{F}>0$ is the output energy and $\sum_{r}\left\langle \dot{n}_{r}\right\rangle \delta\mu_{r}<0$ is an input energy.
For adiabatic quantum pumps, $\sum_{r}\left\langle \dot{n}_{r}\right\rangle \delta\mu_{r}>0$ is the output energy and $\mathcal{W}_{F}<0$ is an input energy. In both cases, $\sum_{r}\left\langle J_{r}\right\rangle \left(\frac{\delta T_{r}}{T_{r}}\right)=0$.
Similarly, for quantum heat engines and quantum heat pumps one obtains
\begin{eqnarray}
1 & \geq & \frac{\mathcal{W}_{F}}{\left(-\sum_{r}\tau\left\langle J_{r}\right\rangle \left(\frac{\delta T_{r}}{T_{r}}\right)-\sum_l \Delta\mathcal{E}_{l}\right)}\qquad\mathrm{and} \notag \\
1 & \geq & \frac{\sum_{r}\tau\left\langle J_{r}\right\rangle \left(\frac{\delta T_{r}}{T_{r}}\right)}{\left(-\mathcal{W}_{F}-\sum_l \Delta\mathcal{E}_{l}\right)}.\label{eq:eta_QHE-QHP}
\end{eqnarray}
For for quantum heat engines, $\mathcal{W}_{F}>0$ is the output energy and $\sum_{r}\left\langle J_{r}\right\rangle \left(\frac{\delta T_{r}}{T_{r}}\right)<0$ is an input energy.
For quantum heat pumps, $\sum_{r}\left\langle J_{r}\right\rangle \left(\frac{\delta T_{r}}{T_{r}}\right)>0$ is the output energy and $\mathcal{W}_{F}<0$ is an input energy. In both cases, $\sum_{r}\left\langle \dot{n}_{r}\right\rangle \delta\mu_{r}=0$.
Note that, except for the term $\sum_l \Delta\mathcal{E}_l$, Eqs. \ref{eq:eta_AQM-AQP} and \ref{eq:eta_QHE-QHP} are equivalent to the known formulas for the efficiency of quantum machines driven only by equilibrium reservoirs~\cite{ludovico2016,ludovico2016B,benenti2017,whitney2018,bustos2019}.
If we neglect this term, seemingly violations of the second law may appear, such as efficiencies of quantum heat engines greater than Carnot's limit (``beyond-Carnot'' efficiencies), or heat spontaneously flowing against the temperature gradient (``break'' of the Clausius inequality).

Here we used the convention that having $J_\alpha>0$, $\dot n_\alpha>0$, or $\dot U_\alpha>0$ means that heat, particles, or energy are entering the reservoir $\alpha$.
Therefore, $\sum_l \Delta\mathcal{E}_l < 0$ means that the NIRs are acting like external power sources.
Note that even when a nonequilibrium reservoir does not exchange energy or particles with the local system, the change of its entropy may still act as a driving force.

Using Eq. \ref{eq:dotS} in Eqs. \ref{eq:eta_AQM-AQP} and \ref{eq:eta_QHE-QHP}, where $\Delta S_l$ is obviously $\tau < \dot S_l >$, provides the upper bound to the efficiency of quantum machines.
It is an upper bound as, of course, other processes can contribute to the global rate of entropy production, e.g., the internal relaxation of the NIR towards its equilibrium. Besides, if the NIR comes from the steady-state of some mesoscopic device, current leakages could increase the global rate of entropy production.
Interestingly, these two phenomena become negligible in the limit $\tau\rightarrow 0$, where the efficiency of the quantum devices should approach Eqs. \ref{eq:eta_AQM-AQP} and \ref{eq:eta_QHE-QHP}.

\section{Landauer-B\"uttiker approach to entropy current}
In the following, we will focus only on the ballistic conduction of electrons in mesoscopic conductors.
In this regime, the particle current of the reservoir $l$ is well described by~\cite{buttiker1985,bode2012}
\begin{eqnarray}
\dot{n}_{l} & = & \frac{1}{h}\sum_{\alpha \neq l}\int_{-\infty}^{\infty}
T_{l,\alpha} \left (f_{\alpha}-f_l\right ) \mathrm{d}\varepsilon, \label{eq:dotn-LB}
\end{eqnarray}
where $h$ the Planck's constant and $T_{\beta,\alpha}$ is the transmittance.
In Eq. \ref{eq:dotn-LB} one can recognize
$\dot{n}_{l}\left(\varepsilon\right) = \frac{1}{h}\sum_{\alpha\neq l}T_{l,\alpha}(f_{\alpha}-f_l)$ as the number of particles per unit energy and unit time entering the reservoir $l$.
Now, inserting $\dot{n}_{l}\left(\varepsilon\right)$ into Eq. \ref{eq:dotS} and integrating over the period $\tau$, gives the Landauer-B\"uttiker expression for the change of the entropy of the reservoir $l$
\begin{eqnarray}
\Delta S_{l} & = & \frac{\tau k_{B}}{h}\sum_{\alpha\neq l }\intop_{-\infty}^{\infty}\left\langle T_{l,\alpha} \right\rangle \left (f_{\alpha}-f_l \right ) \ln\left(\frac{1}{f_{l}} + 1\right)\mathrm{d}\varepsilon.\label{eq:Qd_final}
\end{eqnarray}
If the interaction of the NIR with the local system involves pumped currents, a similar formula can be obtained by using the expressions derived in, e.g., Refs.~\cite{brouwer1998} or~\cite{,bode2012} for $\dot{n}_{l}\left(\varepsilon\right)$.
\begin{figure}
\includegraphics[width=3.0in, trim=0.0in 0.0in 0.0in 0.0in, clip=true]{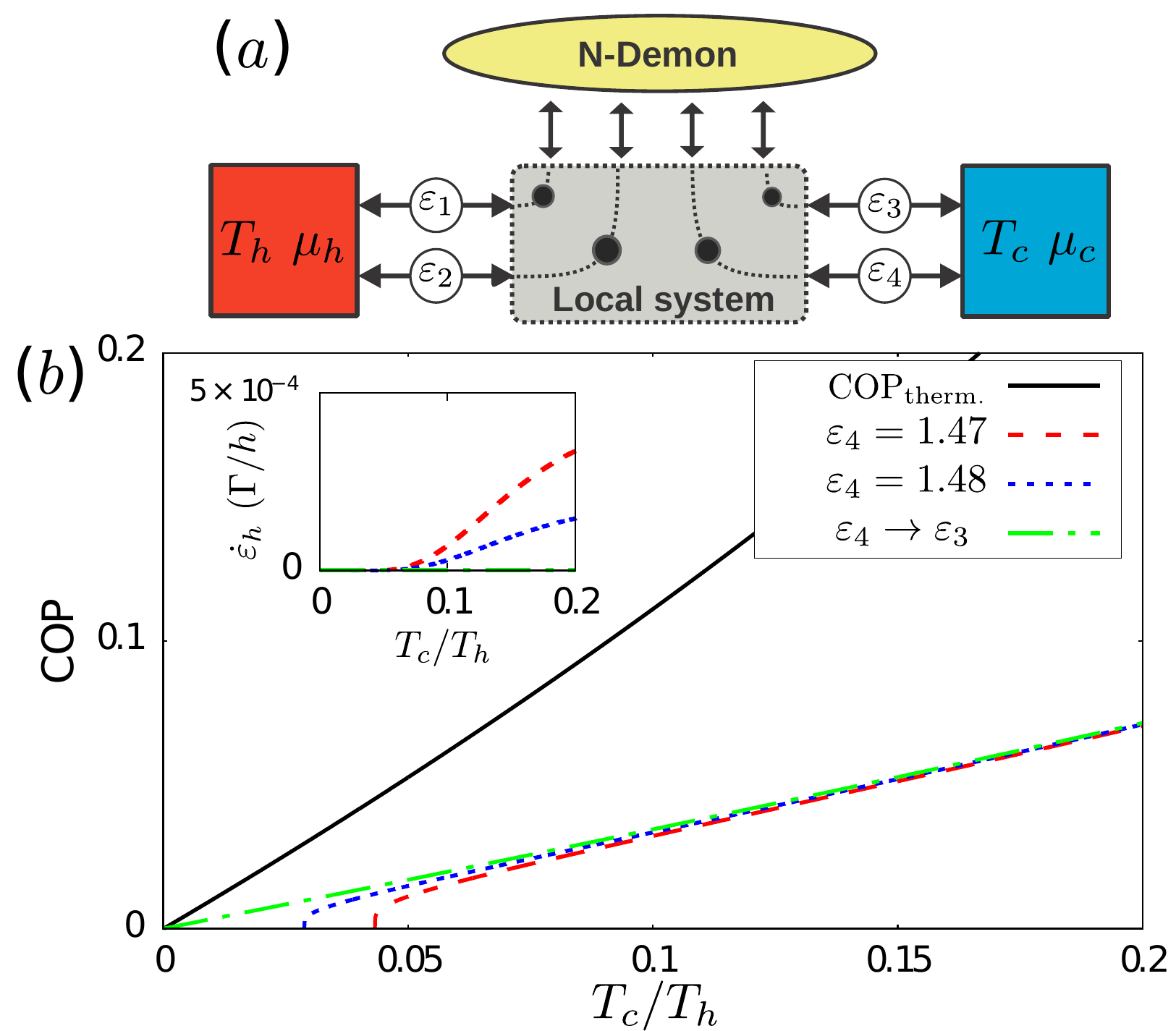}
\caption{
\textbf{(a) - } A local system, consisting of four quantum dots with one narrow resonance each, at energies $\varepsilon_i$, is connected to a NIR (the N demon) and two equilibrium reservoirs, denoted by $h$ and $c$, for hot and cold respectively. Because of the action of the N demon, heat flow ``spontaneously'' from the cold to the hot reservoir.
\textbf{(b) - } Coefficient of performance (COP) of the device depicted in Fig. \ref{fig:2}(a). The inset shows the power of the device ($\dot\varepsilon_h$). The parameters used are $k_B T_h = \mu_h=\mu_c = \varepsilon_1=1$, $\varepsilon_2=2$, and $\varepsilon_3=1.5$}
\label{fig:2}
\end{figure}

\section{Example}
Let us consider a local system coupled to two reservoirs $h$ and $c$ in thermodynamic equilibrium with distribution functions $f_{h}$ and $f_{c}$, temperatures $T_{h}$ and $T_{c}$ (with $T_{h}>T_{c}$), and chemical potentials $\mu_{h}$ and $\mu_{c}$ (with $\mu_{h}=\mu_{c}=\mu_0$).
The local system is also coupled to a third reservoir out of equilibrium with distribution function $f_{l}$, which, given its connection to the local system, works as an intermediary between the equilibrium reservoirs, see Fig. \ref{fig:2}(a).
In the problem we are interested in, the NIR acts as a Maxwell's demon or, more appropriately, as an N demon~\cite{sanchez2019}. The ``demon'' condition implies that the reservoir seemingly ``breaks'' the second law of thermodynamics without exchanging energy or particles with the local system ($\dot{\varepsilon}_{l}=0$ and $\dot{n}_{l}=0$ respectively).
Although the demon condition resembles the voltmeter condition of the fictitious probe model~\cite{buttiker1986,damato1990}, it should not be confused with the measurement-feedback scheme of a standard Maxwell's demon~\cite{koski2014,camati2016,vidrighin2016,chida2017,masuyama2018,strasberg2013}.

Within the Landauer-B\"uttiker approach, the demon condition imposes $\dot{n}_{l}=0$ to Eq. \ref{eq:dotn-LB}, while the condition $\dot{\varepsilon}_{l}=0$ implies
\begin{eqnarray}
\dot{\varepsilon}_{l} & = &\frac{1}{h}\int_{-\infty}^{\infty}\varepsilon
\sum_{r}T_{l,r} \left ( f_{r}-f_l \right ) \mathrm{d}\varepsilon =0.\label{eq:dote_d}
\end{eqnarray}
For simplicity, we assume $T_{r,l}=T_{l,r}$ and that the N demon does not exchange particles in a net way with any of the equilibrium reservoirs ($\dot{n}_{r}=0$), only energy ($\dot{\varepsilon}_{r} \neq 0$).
The problem is to find the distribution function $f_{l}$ (if it exists) such that it produces the non-trivial result $\dot{\varepsilon}_{h}>0$.
To simplify the problem even further, we will consider that $T_{h,l}$ presents two resonances centered at energies $\varepsilon_{1}$ and $\varepsilon_{2}$, where
$\left|\varepsilon_{1}-\varepsilon_{2}\right|$ is much larger than their characteristic width $\Gamma$.
The same is true for $T_{c,l}$, which presents resonances at $\varepsilon_{3}$ and $\varepsilon_{4}$ and where $\left|\varepsilon_{3}-\varepsilon_{4}\right| \gg \Gamma$
Furthermore, we will consider that transmittances are one at their peaks, and that $\Gamma$ is much smaller than the details of the distribution functions $f_{c}$, $f_{h}$, and $f_{l}$.
In this way, the integrals involved in the calculation of all currents (see, e.g., Eqs. \ref{eq:dotn-LB} and \ref{eq:dote_d}) turn into summations where the unknowns are now $f_{l}\left(\varepsilon_{i}\right)$, i.e., the values of $f_{l}$ at energies $\varepsilon_{i}$.
We find that when the local system presents four resonances (four different values of $\varepsilon_{i}$) it is possible to find physical solutions ($0\leq f_{l}(\varepsilon_{i})\leq1$) such that $\dot{\varepsilon}_{h}>0$.
In such a case, see Appendix~\ref{app:example}, the set of equations can be written as:
\begin{eqnarray}
f_{l}\left(\varepsilon_{1,2}\right) & = & f_{h}\left(\varepsilon_{1,2}\right) \mp \left(\frac{\varepsilon_{3}-\varepsilon_{4}}{\varepsilon_{1}-\varepsilon_{2}}\right)\left[f_{c}\left(\varepsilon_{4}\right)-f_{l}\left(\varepsilon_{4}\right)\right] \notag  \\
f_{l}\left(\varepsilon_{3}\right) & = & f_{c}\left(\varepsilon_{3}\right)+f_{c}\left(\varepsilon_{4}\right)-f_{l}\left(\varepsilon_{4}\right).
\label{eq:example}
\end{eqnarray}
This set of three equations is under-determined and thus has infinite solutions. However, we find that the choice $f_{l}(\varepsilon_{4})=-f_{c}(\varepsilon_{3})+2f_{c}(\varepsilon_{4})$ guarantees the desired condition $\dot{\varepsilon}_{h}>0$, where the energy current yields (see Appendix~\ref{app:example})
\begin{eqnarray}
\dot{\varepsilon}_{h} & = & -\frac{\Gamma}{h}\left(\varepsilon_{3}-\varepsilon_{4}\right)\left[f_{c}\left(\varepsilon_{3}\right)-f_{c}\left(\varepsilon_{4}\right)\right].
\label{eq:dote_h}
\end{eqnarray}
Note that not every combination of parameters ($T_{h}$, $T_{c}$, $\mu_{0}$, and $\varepsilon_{i}$) give physical solutions ($0\leq f_{l}(\varepsilon_{i})\leq1$), given our choice of $f_{l}(\varepsilon_{4})$.
If unphysical values of $f_{l}(\varepsilon_{i})$ are found, that means the N demon is unable to pump heat under the studied conditions.

The efficiency of a quantum heat pump is usually discussed in terms of a coefficient of performance (COP)~\cite{benenti2017,whitney2018,bustos2019}.
The value of COP is the ratio of cooling provided to energy required which, according to the discussions after Eqs. \ref{eq:DeltaE} and Eq. \ref{eq:eta_QHE-QHP}, is
\begin{equation}
\mathrm{COP} = \frac{\dot{\varepsilon}_{h}} {(T_{0}\dot{S}_{l})} ,
\end{equation}
or (see Appendix ~\ref{app:example})
\begin{eqnarray}
\mathrm{COP} & & =
\left [
\left (\frac{k_{B}T_{h}}{\varepsilon_{1}-\varepsilon_{2}} \right )
\ln \left( \left[ \frac{1-f_{l}\left(\varepsilon_{1}\right)}{1-f_{l}\left(\varepsilon_{2}\right)}\right] \frac{f_{l}\left(\varepsilon_{2}\right)}{f_{l}\left(\varepsilon_{1}\right)} \right) +  \right . \notag \\
& &
\left .
\left (\frac{k_{B}T_{h}}{\varepsilon_{3}-\varepsilon_{4}} \right )
\ln \left (  \left[\frac{1-f_{l}\left(\varepsilon_{4}\right)}{1-f_{l}\left(\varepsilon_{3}\right)}\right] \frac{f_{l}\left(\varepsilon_{3}\right)}{f_{l}\left(\varepsilon_{4}\right)}
\right)
\right ]^{-1}.
\end{eqnarray}
Here, we make $T_0=T_h$, $\dot{S}_{l}= \Delta S_l /\tau$  and used Eq. \ref{eq:Qd_final}.
Note that, according to Eq. \ref{eq:eta_QHE-QHP} the efficiency of the device is bounded as $0\leq\mathrm{COP}\leq\ T_{c}/(T_{h}-T_{c})$.

In Fig. \ref{fig:2} we show the COP and the power of the quantum heat pump discussed above.
As expected, the efficiencies always lay below the thermodynamic limit $\mathrm{COP}_{\mathrm{therm.}}=T_{c}/(T_{h}-T_{c})$. Moreover, the efficiencies tend to zero when the temperature of the cold reservoir approaches absolute zero, in accordance with the third law of thermodynamics.
The typical power/efficiency trade-off of this kind of machines is also present (compare the central panel of Fig \ref{fig:2}-(b) with its inset).
In addition, it is interesting that there is a minimum temperature, different from zero, below which the cold reservoir cannot be cooled further. This limiting temperature approaches zero only for $\varepsilon_3 \rightarrow \varepsilon_4$ where (see Appendix~\ref{app:example})
\begin{equation}
\mathrm{COP} = \frac{T_c}{3 T_h-T_c} .
\end{equation}

\section{Conclusions}
We provided a general approach for calculating, in a thermodynamically consistent way, the upper bound of the efficiency of NIR-driven quantum machines without relying on any knowledge of how the NIRs were actually made.
This may contribute to the understanding and development of a broader class of quantum machines. In particular, including entropy currents into their analysis, besides energy and particle currents, may be key to shedding light on phenomena such as ``beyond Carnot'' efficiencies or ``breaking'' of Clausius inequalities.

\section{Acknowledgements} We acknowledge discussions with Hernan Calvo and Lucas Fern\'andez-Alc\'azar as well as financial support by
Consejo Nacional de Investigaciones Cient\'ificas y T\'ecnicas (CONICET); Secretar\'ia de Ciencia y Tecnolog\'ia de la Universidad Nacional de C\'ordoba (SECYT-UNC); and Agencia Nacional de Promoción Científica y Tecnológica (ANPCyT, PICT-2018-03587).

\appendix

\section{Entropy current using effective temperatures\label{app:T_eff}}
Suppose one has a reservoir $\alpha$ of $N_{\alpha}$ noninteracting particles, to which we inject a small number of particles.
As we are dealing with noninteracting particles, we can define the contribution to the entropy of each particle $S_{\alpha(1)}$, in terms of the von Neumann entropy of the single-particle density matrix $\rho_{\alpha}$
\begin{eqnarray}
S_{\alpha(1)} & = & -k_{B}\mathrm{Tr}\left(\rho_{\alpha}\ln\rho_{\alpha}\right)
\end{eqnarray}
Deriving $S_{\alpha(1)}$ with respect to time, we obtain
\begin{eqnarray}
\dot{S}_{\alpha(1)} & = & -k_{B}\sum_{i}\dot{\rho}_{\alpha i}\ln\rho_{\alpha i}.\label{eq:dotS-2}
\end{eqnarray}
Above, we used $\sum_{i}\dot{\rho}_{\alpha i}=0$, and write the density matrix $\rho_{\alpha}$ in the energy basis, which we assume diagonalizes it. 
The reservoir $\alpha$ is not necessarily in equilibrium but, for the sake of convenience, we are going to take the following generic functional form for $\rho_{\alpha i}$, or the probability of finding the particle in the eigenstate $i$ of the Hamiltonian of the reservoir
$\alpha$, 
\begin{eqnarray}
\rho_{\alpha i} & \equiv & \frac{\exp\left[-\beta_{\alpha}\left(\epsilon_{i}\right)\left(\epsilon_{i}-\epsilon_{\alpha F}\right)\right]}{Z_{\alpha}},
\end{eqnarray}
where $\epsilon_{i}$ is the $i$-th eigenenergy of the single-particle
Hamiltonian of the reservoir, $\beta_{\alpha}\left(\epsilon_{i}\right)$ is not a constant but just some arbitrary function of $\epsilon_{i}$, similarly, $\epsilon_{\alpha F}$ is an arbitrary constant not necessarily with physical meaning, and $Z_{\alpha}$ is the normalization constant.
Using this in Eq. \ref{eq:dotS-2} yields
\begin{eqnarray}
\dot{S}_{\alpha(1)} & = & k_{B}\sum_{i}\dot{\rho}_{\alpha i}\beta_{\alpha}\left(\epsilon_{i}\right)\left(\epsilon_{i}-\epsilon_{\alpha F}\right)
\end{eqnarray}
where we again used $\sum_{i}\dot{\rho}_{\alpha i}=0$. 
Now we wonder how $\rho_{\alpha i}$ is related to $n_{\alpha i}$ (the number of particles in the reservoir with energy within the interval $\delta\epsilon$ around $\epsilon_{i}$).
This is given by
\begin{eqnarray}
\rho_{\alpha i} & \approx & \frac{n_{\alpha i}}{N_{\alpha}}.
\end{eqnarray}
Above we assumed, $\delta\epsilon$ is small enough such as $\rho_{\alpha}(\epsilon_{i}-\delta\epsilon/2)\approx\rho_{\alpha}(\epsilon_{i}+\delta\epsilon/2)$, and $N_{\alpha}$ is large enough such as the statistical error implicit in the equation is negligible. 
Deriving $\rho_{\alpha i}$ with respect to time gives
\begin{eqnarray}
\dot{\rho}_{\alpha i} & \approx & \frac{\left[\dot{n}_{\alpha i}-n_{\alpha i}\frac{\dot{N}_{\alpha}}{N_{\alpha}}\right]}{N_{\alpha}}
\end{eqnarray}
We are interested in the case $\frac{\dot{N}_{\alpha}}{N_{\alpha}}\approx0$, which corresponds to the limit of a large reservoir. Using this we obtain
\begin{eqnarray}
\dot{S}_{\alpha} & = & k_{B}\sum_{i}\dot{n}_{\alpha i}\beta_{\alpha}\left(\epsilon_{i}\right)\left(\epsilon_{i}-\epsilon_{\alpha F}\right)
\end{eqnarray}
where $\dot{S}_{\alpha}=N_{\alpha}\dot{S}_{\alpha(1)}$ is the total change of the entropy of the reservoir, which we assumed is composed of $N_{\alpha}$ noninteracting particles. 
Finally, defining the particle's current density $\dot{n}_{\alpha}\left(\epsilon_{i}\right)=\dot{n}_{\alpha i}/\delta\epsilon$ and turning the summation into an integral we get
\begin{eqnarray}
\dot{S}_{\alpha} & = & \intop\dot{n}_{\alpha}\left(\epsilon\right)
\frac{\left(\epsilon-\epsilon_{\alpha F}\right)}{T_{\mathrm{eff}}\left(\epsilon\right)}
\mathrm{d}\epsilon ,
\end{eqnarray}
where $T_{\mathrm{eff}}\left(\epsilon\right)=1/k_{B}\beta_{\alpha}\left(\epsilon\right)$ is the energy-dependent effective temperature.
If we compare this formula with Eq. \ref{eq:dotS} of the main text, we conclude that $T_{\mathrm{eff}}\left(\epsilon\right)$ is given by
\begin{eqnarray}
T_{\mathrm{eff}}\left(\epsilon\right) & = & \frac{\left(\epsilon-\epsilon_{\alpha F}\right)}{k_{B}\ln\left(\frac{1}{f_{\alpha}\left(\epsilon\right)}\pm1\right)},
\end{eqnarray}
where the $+$ sign is for bosons and the $-$ sign is for fermions. This expression for the effective temperature is the same as that derived in Ref.~\cite{alicki2015} for harmonic-oscillators nonequilibrium-baths.

\section{Interpretation of the term $\Delta\mathcal{E}$\label{app:DE}}

We start from the expression for $\Delta\mathcal{E}_l$, Eq. \ref{eq:DeltaE} of the main text,
\begin{eqnarray}
\Delta\mathcal{E}_l & = & \Delta U_l-\mu_{0}\Delta n_l-T_{0}\Delta S_l.
\end{eqnarray}
Now, let us impose arbitrary values of temperature $T_l$ and chemical potential $\mu_l$ to the $l$-reservoir so that $\Delta U_l\equiv T_l\Delta S_l^{\mathrm{eq}}+\Delta n_l\mu_l$, where $\Delta S_l^{\mathrm{eq}}\equiv(\Delta U_l-\Delta n_l\mu_l)/T_l$.
Using this, the above equation can be rewritten as
\begin{eqnarray}
\Delta\mathcal{E}_l & = & T_l\Delta S_l^{\mathrm{eq}}+\mu_l\Delta n_l-\mu_{0}\Delta n_l-T_{0}\Delta S_l\nonumber \\
 & = & \Delta n_l\delta\mu_l+\Delta S_l^{\mathrm{eq}}\delta T_l+T_{0}\left(\Delta S_l^{\mathrm{eq}}-\Delta S_l\right)\label{eq:DE_l-02}
\end{eqnarray}
where we used $\delta T_l=T_l-T_{0}$ and $\delta\mu_l=\mu_l-\mu_{0}$.
The terms $\Delta n_l\delta\mu_l$ and $\Delta S_l^{\mathrm{eq}}\delta T_l$ of the right-hand side of the equation correspond to the change of the energy of the ``equilibrium'' $l$-reservoir with temperature $T_l$ and chemical potential $\mu_l$.
The change of the entropy of the third term can be rewritten as
\begin{eqnarray}
-\left(\Delta S_l^{\mathrm{eq}}-\Delta S_l\right) & = & 
\left(S_{l-\mathrm{ini}}^{\mathrm{eq}}-S_{l-\mathrm{ini}}\right)
\notag \\ & &
+\left(S_{l-\mathrm{end}}-S_{l-\mathrm{end}}^{\mathrm{eq}}\right)
\end{eqnarray}
The right-hand side of the above equation yields the change of the entropy of the $l$-reservoir when it is transformed from the initial nonequilibrium state to the initial equilibrium state, plus the change of the entropy of the $l$-reservoir from the final equilibrium state to the final nonequilibrium state.
Now, as the total change of the entropy of the universe is zero during the whole process then, the change of the entropy of the $l$-reservoir is equal in magnitude but opposite in sign to the change of the entropy of the environment, assumed in equilibrium at temperature $T_{0}$.
Therefore, the term $T_{0}\left(\Delta S_l^{\mathrm{eq}}-\Delta S_l\right)$ can be interpreted as the heat absorbed by the environment during the process.

\section{N demon example\label{app:example}}

Let us consider a local system coupled to two reservoirs in thermodynamic equilibrium at temperatures $T_{h}$ and $T_{c}$ (where $T_{h}>T_{c}$) and chemical potentials $\mu_{h}$ and $\mu_{c}$ (where $\mu_{h}=\mu_{c}=\mu_{0}$).
Each of the reservoirs has associated a Fermi-Dirac distribution function
$f_{r}(\varepsilon)=\left(\exp\left[(\varepsilon-\mu_{r})/(k_{B}T_{r})\right]+1\right)^{-1}$,
where $r=h,c$.
We will introduce a third reservoir out of equilibrium with a distribution function $f_l(\varepsilon)$, which represents the N demon. Due to the configuration of the local system, the N demon acts as an intermediary between the reservoirs in thermodynamic equilibrium through two resonances for each reservoir, see Fig. 2 (a) in the main text.
With this configuration, and within the Landauer-B\"uttiker approach of quantum transport, the particle and energy currents ($\dot{n}$ and $\dot{\varepsilon}$ respectively) are:
\begin{eqnarray}
\dot{n}_{r} & = &\frac{1}{h}\int^{\infty}_{-\infty} T_{r,l}\left[f_l\left(\varepsilon\right) -    
 f_{r}\left(\varepsilon\right)\right]\mathrm{d}\varepsilon
\nonumber \\ 
\dot{\varepsilon}_{r} & = & \frac{1}{h}\int^{\infty}_{-\infty}\varepsilon T_{r,l}\left[f_l\left(\varepsilon\right)-f_{r}\left(\varepsilon\right)\right]\mathrm{d}\varepsilon
\nonumber \\
\dot{n}_l & = & \frac{1}{h}\int^{\infty}_{-\infty} \sum_{r}T_{l,r}\left[f_{r}\left(\varepsilon\right)-f_l\left(\varepsilon\right)\right]\mathrm{d}\varepsilon
 \nonumber \\
\dot{\varepsilon}_l & = &\frac{1}{h}\int^{\infty}_{-\infty} \varepsilon\sum_{r}T_{l,r}\left[f_{r}\left(\varepsilon\right)-f_l\left(\varepsilon\right)\right]\mathrm{d}\varepsilon \label{eq:LB_n_e}
\end{eqnarray}
where $h$ is the Planck's constant, index $r$ is $\{h,c\}$, $T_{r,l}$ is the transmittance between the $r$ and $l$ reservoirs (we are assuming $T_{r,l}=T_{l,r}$).
Note that the above equations naturally fulfill particle and energy conservation laws ($\dot{n}_{h}+\dot{n}_{c}=-\dot{n}_l$ and $\dot{\varepsilon}_{h}+\dot{\varepsilon}_{c}=-\dot{\varepsilon}_l$ respectively).

The demon condition imposes the requirements $\dot{n}_l=0$ and $\dot{\varepsilon}_l=0$ to Eqs. \ref{eq:LB_n_e}.
To simplify the problem, we also added the conditions $\dot{n}_{r}=0$ and adopted the following simple functional form for the transmittances
\begin{eqnarray}
T_{h,l} & = & \sum_{i=1,2}\Theta\left[\varepsilon-\left(\varepsilon_{i}-\frac{\Gamma}{2}\right)\right]-\Theta\left[\varepsilon-\left(\varepsilon_{i}+\frac{\Gamma}{2}\right)\right]
\quad \mathrm{and} \nonumber \\
T_{c,l} & = & \sum_{i=3,4}\Theta\left[\varepsilon-\left(\varepsilon_{i}-\frac{\Gamma}{2}\right)\right]-\Theta\left[\varepsilon-\left(\varepsilon_{i}+\frac{\Gamma}{2}\right)\right],
\end{eqnarray}
where $\Theta\left(\varepsilon\right)$ is the Heaviside step function, $\varepsilon_{i}$ is the center of the resonance $i$, and $\Gamma$ is the characteristic width of the resonances.
Finally, we assumed $\Gamma$ is sufficiently small such as $\Gamma < |\varepsilon_{1}-\varepsilon_{2}|$,
$\Gamma<|\varepsilon_{3}-\varepsilon_{4}|$, and the distribution functions {[}$f_{c}(\varepsilon)$, $f_{h}(\varepsilon)$, and $f_{l}(\varepsilon)${]} can be considered constant within the energy intervals $(\varepsilon_{i}-\Gamma/2) \leq \varepsilon \leq (\varepsilon_{i}+\Gamma/2)$.
Using all these into Eq. \ref{eq:LB_n_e} gives
\begin{eqnarray}
0 & = & f_l\left(\varepsilon_{1}\right)-f_{h}\left(\varepsilon_{1}\right)+f_l\left(\varepsilon_{2}\right)-f_{h}\left(\varepsilon_{2}\right)
\nonumber \\
0 & = & f_l\left(\varepsilon_{3}\right)-f_{c}\left(\varepsilon_{3}\right)+f_l\left(\varepsilon_{4}\right)-f_{c}\left(\varepsilon_{4}\right)\
\nonumber \\
0 & = & 
 \varepsilon_{1}\left[f_l\left(\varepsilon_{1}\right)-f_{h}\left(\varepsilon_{1}\right)\right]
+\varepsilon_{2}\left[f_l\left(\varepsilon_{2}\right)-f_{h}\left(\varepsilon_{2}\right)\right] 
\nonumber \\ & &
+\varepsilon_{3}\left[f_l\left(\varepsilon_{3}\right)-f_{c}\left(\varepsilon_{3}\right)\right]
+\varepsilon_{4}\left[f_l\left(\varepsilon_{4}\right)-f_{c}\left(\varepsilon_{4}\right)\right]
\end{eqnarray}
where we also used particle and energy conservation.
After some simple algebraic manipulations, the above equations turn into Eq. \ref{eq:example} of the main text, i.e.,
\begin{eqnarray}
f_l\left(\varepsilon_{1}\right) & = & f_{h}\left(\varepsilon_{1}\right)-\left(\frac{\varepsilon_{3}-\varepsilon_{4}}{\varepsilon_{1}-\varepsilon_{2}}\right)\left[f_{c}\left(\varepsilon_{4}\right)-f_l\left(\varepsilon_{4}\right)\right]
\nonumber \\
f_l\left(\varepsilon_{2}\right) & = & f_{h}\left(\varepsilon_{2}\right)+\left(\frac{\varepsilon_{3}-\varepsilon_{4}}{\varepsilon_{1}-\varepsilon_{2}}\right)\left[f_{c}\left(\varepsilon_{4}\right)-f_l\left(\varepsilon_{4}\right)\right]\label{eq:17_main_text}
\nonumber \\
f_l\left(\varepsilon_{3}\right) & = & f_{c}\left(\varepsilon_{3}\right)+f_{c}\left(\varepsilon_{4}\right)-f_l\left(\varepsilon_{4}\right)
\nonumber \\
f_l\left(\varepsilon_{4}\right) & = & -f_{c}\left(\varepsilon_{3}\right)+2f_{c}\left(\varepsilon_{4}\right).
\end{eqnarray}
Note that we included above our choice for $f_l\left(\varepsilon_{4}\right)$.

\subsection{Power}

Within the Landauer-B\"uttiker approach, the power of the refrigerator in the example is
\begin{eqnarray}
\dot{\varepsilon}_{h} & = & \frac{\Gamma}{h}\left(\varepsilon_{1}\left[f_l\left(\varepsilon_{1}\right)-f_{h}\left(\varepsilon_{1}\right)\right]+\varepsilon_{2}\left[f_l\left(\varepsilon_{2}\right)-f_{h}\left(\varepsilon_{2}\right)\right]\right)
\end{eqnarray}
If we insert Eqs. \ref{eq:17_main_text} into the above equation we obtain Eq. \ref{eq:dote_h} of the main text, i.e.,
\begin{eqnarray}
\dot{\varepsilon}_{h} & = & -\frac{\Gamma}{h}\left(\varepsilon_{3}-\varepsilon_{4}\right)\left[f_{c}\left(\varepsilon_{3}\right)-f_{c}\left(\varepsilon_{4}\right)\right].\label{eq:dote_h_model}
\end{eqnarray}
Since $f_{c}\left(\varepsilon\right)$ is an equilibrium distribution function: if $\varepsilon_{3}>\varepsilon_{4}$ then $f_{c}\left(\varepsilon_{3}\right)<f_{c}\left(\varepsilon_{4}\right)$,
while if $\varepsilon_{3}<\varepsilon_{4}$ then $f_{c}\left(\varepsilon_{3}\right)>f_{c}\left(\varepsilon_{4}\right)$.
Therefore, the desired condition, heat flowing against the temperature gradient ($\dot{\varepsilon}_{h}\geq0$), is guaranteed.

\subsection{COP}

As discussed in the main text the coefficient of performance (COP) in the example is
\begin{eqnarray}
\mathrm{COP} & = & \frac{\dot{\varepsilon}_{h}}{T_{0}\dot{S}_l}
\end{eqnarray}
The expression for $\dot{\varepsilon}_{h}$ is given in Eq. \ref{eq:dote_h_model}, while the expression for $\dot{S}_l$ can be derived by using Eq. \ref{eq:Qd_final} of the main text on the example treated here. That
gives
\begin{widetext}
\begin{eqnarray}
\dot{S}_l & = & \frac{k_{B}}{h}\sum_{r} \int^{-\infty}_{\infty} \left(T_{l,r}\left[f_{r}\left(\varepsilon\right)-f_l\left(\varepsilon\right)\right]\mathrm{d}\varepsilon\right)\ln\left(\frac{1}{f_l\left(\varepsilon\right)}-1\right)d\varepsilon
 \nonumber \\
 &  & \simeq\frac{\Gamma k_{B}}{h}\left(\left[f_{h}\left(\varepsilon_{1}\right)-f_l\left(\varepsilon_{1}\right)\right]\ln\left(\frac{1}{f_l\left(\varepsilon_{1}\right)}-1\right)+\left[f_{h}\left(\varepsilon_{2}\right)-f_l\left(\varepsilon_{2}\right)\right]\ln\left(\frac{1}{f_l\left(\varepsilon_{2}\right)}-1\right)\right.+
 \nonumber  \\
 &  & +\left.\left[f_{c}\left(\varepsilon_{3}\right)-f_l\left(\varepsilon_{3}\right)\right]\ln\left(\frac{1}{f_l\left(\varepsilon_{3}\right)}-1\right)+\left[f_{c}\left(\varepsilon_{4}\right)-f_l\left(\varepsilon_{4}\right)\right]\ln\left(\frac{1}{f_l\left(\varepsilon_{4}\right)}-1\right)\right].
\end{eqnarray}
Now, inserting Eqs. \ref{eq:17_main_text} and \ref{eq:dote_h_model} into the above expression,
and using $\frac{1}{f_l\left(\varepsilon_{1}\right)}-1=\frac{1-f_l\left(\varepsilon_{1}\right)}{f_l\left(\varepsilon_{1}\right)}$ yields
\begin{eqnarray}
\dot{S}_l & \simeq & \frac{\Gamma k_{B}}{h}\left(\left(\frac{\varepsilon_{3}-\varepsilon_{4}}{\varepsilon_{1}-\varepsilon_{2}}\right)\left[f_{c}\left(\varepsilon_{4}\right)-f_l\left(\varepsilon_{4}\right)\right]\ln\left(\frac{1}{f_l\left(\varepsilon_{1}\right)}-1\right)-\left(\frac{\varepsilon_{3}-\varepsilon_{4}}{\varepsilon_{1}-\varepsilon_{2}}\right)\left[f_{c}\left(\varepsilon_{4}\right)-f_l\left(\varepsilon_{4}\right)\right]\ln\left(\frac{1}{f_l\left(\varepsilon_{2}\right)}-1\right)\right.
 \nonumber \\
 &  & \left.-\left[f_{c}\left(\varepsilon_{4}\right)-f_l\left(\varepsilon_{4}\right)\right]\ln\left(\frac{1}{f_l\left(\varepsilon_{3}\right)}-1\right)+\left[f_{c}\left(\varepsilon_{4}\right)-f_l\left(\varepsilon_{4}\right)\right]\ln\left(\frac{1}{f_l\left(\varepsilon_{4}\right)}-1\right)\right)
 \nonumber \\
 & \simeq & -\dot{\varepsilon}_{h} k_B \left(\frac{1}{\varepsilon_{1}-\varepsilon_{2}}\left[\ln\left(\frac{1}{f_l\left(\varepsilon_{1}\right)}-1\right)-\ln\left(\frac{1}{f_l\left(\varepsilon_{2}\right)}-1\right)\right]+\frac{1}{\varepsilon_{3}-\varepsilon_{4}}\left[\ln\left(\frac{1}{f_l\left(\varepsilon_{4}\right)}-1\right)-\ln\left(\frac{1}{f_l\left(\varepsilon_{3}\right)}-1\right)\right]\right).
 \nonumber \\
 & \simeq & -\dot{\varepsilon}_{h} k_B \left(\frac{\ln\left[\left(\frac{1-f_l\left(\varepsilon_{1}\right)}{1-f_l\left(\varepsilon_{2}\right)}\right)\frac{f_l\left(\varepsilon_{2}\right)}{f_l\left(\varepsilon_{1}\right)}\right]}{\left(\varepsilon_{1}-\varepsilon_{2}\right)}+\frac{\ln\left[\left(\frac{1-f_l\left(\varepsilon_{4}\right)}{1-f_l\left(\varepsilon_{3}\right)}\right)\frac{f_l\left(\varepsilon_{3}\right)}{f_l\left(\varepsilon_{4}\right)}\right]}{\left(\varepsilon_{3}-\varepsilon_{4}\right)}\right)
\end{eqnarray}
Using the above expression we obtain the final result for the COP,
\begin{eqnarray}
\mathrm{COP} & = & \left[\left(\frac{k_{B}T_{0}}{\varepsilon_{1}-\varepsilon_{2}}\right)\ln\left(\left[\frac{1-f_l\left(\varepsilon_{2}\right)}{1-f_l\left(\varepsilon_{1}\right)}\right]\frac{f_l\left(\varepsilon_{1}\right)}{f_l\left(\varepsilon_{2}\right)}\right)+\left(\frac{k_{B}T_{0}}{\varepsilon_{3}-\varepsilon_{4}}\right)\ln\left(\left[\frac{1-f_l\left(\varepsilon_{3}\right)}{1-f_l\left(\varepsilon_{4}\right)}\right]\frac{f_l\left(\varepsilon_{4}\right)}{f_l\left(\varepsilon_{3}\right)}\right)\right]^{-1}.
\label{eq:COP}
\end{eqnarray}
\end{widetext}

\subsection{COP for $\lim_{\varepsilon_{4}\rightarrow\varepsilon_{3}}$}

We start by rewriting the logarithmic functions of Eq. \ref{eq:COP} as
\begin{widetext}
\begin{eqnarray}
\ln\left[\left(\frac{1-f_l\left(\varepsilon_{1}\right)}{1-f_l\left(\varepsilon_{2}\right)}\right)\frac{f_l\left(\varepsilon_{2}\right)}{f_l\left(\varepsilon_{1}\right)}\right] & = & 2\textrm{\textrm{arctanh\ensuremath{\left(\frac{\tanh\left(\frac{\varepsilon_{1}-\varepsilon_{2}}{2k_{B}T_{h}}\right)\left\{ 1-2\frac{A_{c}}{A_{h}}\right\} }{\frac{\tanh\left(\frac{\varepsilon_{1}-\varepsilon_{2}}{k_{B}T_{h}}\right)}{\left(\frac{\varepsilon_{1}-\varepsilon_{2}}{k_{B}T_{h}}\right)}A_{c}\left\{ \frac{A_{c}}{A_{h}}-1\right\} +1}\right)}}}
 \nonumber \\
\ln\left[\left(\frac{1-f_l\left(\varepsilon_{4}\right)}{1-f_l\left(\varepsilon_{3}\right)}\right)\frac{f_l\left(\varepsilon_{3}\right)}{f_l\left(\varepsilon_{4}\right)}\right] & = & 2\textrm{\textrm{arctanh\ensuremath{\left(\frac{-3\tanh\left(\frac{\varepsilon_{3}-\varepsilon_{4}}{2k_{B}T_{c}}\right)}{2\frac{\tanh\left(\frac{\varepsilon_{3}-\varepsilon_{4}}{2k_{B}T_{c}}\right)}{\left(\frac{\varepsilon_{3}-\varepsilon_{4}}{2k_{B}T_{c}}\right)}A_{c}+1}\right)}}}
\label{eq:lim_ln}
\end{eqnarray}
\end{widetext}
where we used Eqs. \ref{eq:17_main_text}, the relation
\begin{equation}
\tanh\left(\frac{\varepsilon-\mu_{0}}{2k_{B}T_{c,h}}\right)=1-2f_{c,h}\left(\varepsilon\right) , 
\end{equation}
and we defined
the auxiliary functions 
\begin{equation}
A_{c}=\left(\frac{\varepsilon_{3}-\varepsilon_{4}}{2k_{B}T_{c}}\right)\left[\tanh\left(\frac{\varepsilon_{3}-\mu_{0}}{2k_{B}T_{c}}\right)-\tanh\left(\frac{\varepsilon_{4}-\mu_{0}}{2k_{B}T_{c}}\right)\right]
\end{equation}
and 
\begin{equation}
A_{h}=\left(\frac{\varepsilon_{1}-\varepsilon_{2}}{2k_{B}T_{h}}\right)\left[\tanh\left(\frac{\varepsilon_{1}-\mu_{0}}{2k_{B}T_{h}}\right)-\tanh\left(\frac{\varepsilon_{2}-\mu_{0}}{2k_{B}T_{h}}\right)\right] ,
\end{equation}
just to make the formulas more compact.

Before analyzing the limit we are interested in, $\lim_{\Delta\rightarrow0}\mathrm{COP}$ where $\Delta=(\varepsilon_{3}-\varepsilon_{4})/(2k_{B}T_{c})$, we will consider the following asymptotic behaviors:
\begin{eqnarray}
\lim_{\Delta\rightarrow0}\left[\tanh\left(\frac{\varepsilon_{3}-\varepsilon_{4}}{2k_{B}T_{c}}\right)\right] & \asymp & \left(\frac{\varepsilon_{3}-\varepsilon_{4}}{2k_{B}T_{c}}\right)
\notag \\
\mathrm{and} & &
\notag \\
\quad\lim_{\Delta\rightarrow0}A_{c} & \asymp 0.
\end{eqnarray}
Using the above into Eqs. \ref{eq:lim_ln} we arrive to
\begin{eqnarray}
{\lim_{\Delta\rightarrow0}\left[\frac{1}{\varepsilon_{1}-\varepsilon_{2}}\ln\left(\left(\frac{1-f_{l}\left(\varepsilon_{1}\right)} {1-f_{l}\left(\varepsilon_{2}\right)}\right)\frac{f_{l}\left(\varepsilon_{2}\right)}{f_{l}\left(\varepsilon_{1}\right)}\right)\right]}
 &\asymp& \nonumber \\
{\lim_{\Delta\rightarrow0}\left[\frac{1}{\varepsilon_{1}-\varepsilon_{2}}2\textrm{\textrm{arctanh\ensuremath{\left(\tanh\left(\frac{\varepsilon_{1}-\varepsilon_{2}}{2k_{B}T_{h}}\right)\right)}}}\right]}
 &\asymp& \nonumber \\
  \frac{1}{k_{B}T_{h}}.
\end{eqnarray}
and
\begin{eqnarray}
\lim_{\Delta\rightarrow0}\left[\frac{1}{\varepsilon_{3}-\varepsilon_{4}}\ln\left(\left(\frac{1-f_l\left(\varepsilon_{4}\right)}{1-f_l\left(\varepsilon_{3}\right)}\right)\frac{f_l\left(\varepsilon_{3}\right)}{f_l\left(\varepsilon_{4}\right)}\right)\right]
 &\asymp& \nonumber \\
\lim_{\Delta\rightarrow0}\left[\frac{1}{\varepsilon_{3}-\varepsilon_{4}}2\textrm{\textrm{arctanh\ensuremath{\left(-3\tanh\left(\frac{\varepsilon_{3}-\varepsilon_{4}}{2k_{B}T_{c}}\right)\right)}}}\right].
\end{eqnarray}
Finally, with the aid of the expression
\begin{eqnarray}
2\textrm{\textrm{arctanh\ensuremath{\left(-3\tanh\left(x\right)\right)}}} & = & -6x+O(x^{3}),
\end{eqnarray}
we find
\begin{eqnarray}
\lim_{\Delta\rightarrow0}\left[\frac{1}{\varepsilon_{3}-\varepsilon_{4}}\ln\left(\left(\frac{1-f_l\left(\varepsilon_{3}\right)}{1-f_l\left(\varepsilon_{4}\right)}\right)\frac{f_l\left(\varepsilon_{4}\right)}{f_l\left(\varepsilon_{3}\right)}\right)\right]
 &\asymp& \nonumber \\
\frac{-3}{k_{B}T_{c}}.
\end{eqnarray}

Now we are in condition of evaluating the asymptotic behavior of COP for $\varepsilon_{3}\rightarrow\varepsilon_{4}$ which, assuming $T_{0}=T_{h}$ and using the above results, gives
\begin{eqnarray}
\lim_{\Delta\rightarrow0}\left[\mathrm{COP}\right] & \asymp & \left(\frac{T_{c}}{3T_{h}-T_{c}}\right)
\end{eqnarray}

\bibliography{DBM-PRB}

\end{document}